\documentclass[aps,showpacs,twocolumn,prl,superscriptaddress]{revtex4}
\usepackage{upgreek, graphicx, subfigure, amsmath}
\newcommand{\eref}[1]{Eq.~(\ref{#1})}
\newcommand{\erefs}[1]{Eqs.~(\ref{#1})}
\newcommand{\fref}[1]{Fig.~\ref{#1}}

\newcommand{\ie}{i.e.}
\newcommand{\rmd}{~\text{d}}
\newcommand{\force}{\boldsymbol{F}}
\newcommand{\im}[1]{\,\text{Im}\!\left\{#1\right\}}
\newcommand{\re}[1]{\,\text{Re}\!\left\{#1\right\}}
\DeclareMathOperator{\Lapprox}{\approx}
\DeclareMathOperator{\Llt}{<}
\makeatletter
\newlength \figwidth
\makeatother
\begin{document}
\title{Optomechanical cooling  with generalized interferometers}
\author{Andr\'e Xuereb}
\email[Corresponding author. Electronic address:~]{andre.xuereb@soton.ac.uk}
\affiliation{School of Physics and Astronomy, University of Southampton, Southampton SO17~1BJ, United Kingdom}
\author{Tim Freegarde}
\affiliation{School of Physics and Astronomy, University of Southampton, Southampton SO17~1BJ, United Kingdom}
\author{Peter Horak}
\affiliation{Optoelectronics Research Centre, University of Southampton, Southampton SO17~1BJ, United Kingdom}
\author{Peter Domokos}
\affiliation{Research Institute of Solid State Physics and Optics, H-1525 Budapest P.O. Box 49, Hungary}
\date{\today}
\begin{abstract}
The fields in multiple-pass interferometers, such as {the Fabry--P\'erot} cavity, exhibit great sensitivity not only to the {presence but} also to the {\emph{motion} of} any scattering object within the optical path. We consider {the general} case of an interferometer comprising an arbitrary configuration of generic `beam {splitters' and} calculate the velocity-dependent radiation field and the light force exerted on a moving scatterer. {We find that a simple} configuration, in which the scatterer interacts with an optical resonator from which it is spatially separated, can enhance the optomechanical friction by several orders of magnitude.
\end{abstract}
\maketitle
Optomechanics~\cite{Marquardt2009} is a rapidly growing field addressing the manipulation of macroscopic scatterers by making use of the mechanical effects of light. The ponderomotive force exhibits a velocity-dependent character which stems from {any retardation of the electromagnetic field present} in such systems. With an appropriate choice of parameters, velocity-dependent terms in the force may lead to viscous damping of motion {\cite{Kippenberg2007}}.
\par
{Pure} Doppler frequency shifting results in a velocity dependent force with a relative magnitude of order $v/c$, which is {generally small} at room temperature {or} below. The laser cooling of atoms, for example, produces a significant cooling effect because it is resonantly enhanced by the atom with the $Q$-factor $\omega/\gamma$ characteristic of an atomic transition ($\omega$ is the frequency of the radiation, $\gamma$ is the linewidth of the transition). This situation can be mimicked in the case of {a moving micro-mirror}, as was proposed in Ref.~\cite{Karrai2008}, {whereby a} photonic crystal having {a} steep frequency-dependent reflection coefficient is mounted {upon} it. In the more general case of a non-resonant scatterer, the sensitivity of the radiation force to the velocity can be enhanced by coupling the moving object to a resonant optical element. This is the case, for example, in several recent optomechanical cooling experiments~\cite{Metzger2004,Arcizet2006,Gigan2006,Schliesser2008}: the thermal vibration of one of the micro-mirrors making up a Fabry--P\'erot-type resonator can be quenched through the radiation pressure of the light field enclosed in the resonator. Several factors limit the efficiency of this mechanism in practice, including the quality of the micromirrors that can be fabricated and the precision with which the cavities can be aligned.
\par
\begin{figure}
\centering
\includegraphics[width=1.2\figwidth]{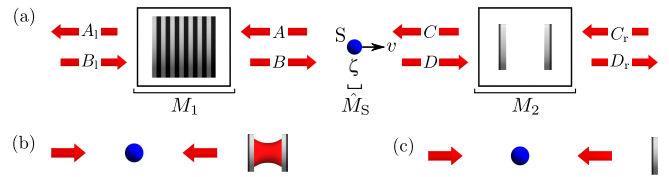}
\caption{(a) The general system consisting of a mobile scatterer, S, between two sets of generic immobile optical elements (we show a Bragg reflector, on the left, and a Fabry--P\'erot-type cavity, on the right, as an example). The mobile scatterer can, \emph{a priori}, represent anything, e.g., an atom or a mirror. We discuss two specific configurations in this Letter: an atom in front of (b) a two-mirror cavity, and (c) a plane mirror~\cite{Xuereb2009a}.}
\label{fig:Models}
\end{figure}

In this Letter we generalize the conventional optomechanical cooling scheme \cite{Metzger2004,Arcizet2006,Gigan2006,Schliesser2008} and calculate the linear response of the electromagnetic field to the motion of an {arbitrary scatterer within a general 1D} configuration of immobile optical elements on either side of it {(see~\fref{fig:Models}(a))}. We find that the field interference can be {significantly} sensitive to motion even if the scatterer {lacks a} specific frequency-dependent reflectivity.
\par
In the second part of this Letter, the role of interference in enhancing the viscous cooling force is analyzed for a simple geometry, in which the scatterer {lies in front of, \emph{but not within}, a standard two-mirror resonator, as in~\fref{fig:Models}(b). With this scheme, which we label `external cavity cooling', one can benefit from the high finesse of the cavity even if the moving object has a low reflectivity.} We thus propose a very general, efficient optomechanical cooling mechanism applicable to a wide class of micro- or mesoscopic {objects.}
\par
We {begin by presenting} the formal solution of the scattering model which we constructed in a recent paper for dealing with a general configuration of one-dimensional optomechanical systems~\cite{Xuereb2009b}. Each element of the system is described by a transfer matrix which relates linearly the field amplitudes on its left-hand side to those on the right-hand side. Transfer matrices for moving scatterers up to linear order in $v/c$ have been constructed.
The transfer matrix of an arbitrary configuration of optical elements is then obtained by matrix multiplication. {A difficulty in analyzing} complex networks originates from the Doppler shift operator $\hat{P}_v$~\cite{Xuereb2009b}, which appears in the transfer matrix of the moving scatterer and acts in the space of the wave vectors rather than in the space of amplitudes: $\hat{P}_vf(k)=f(k+k_0v/c)$, for any function $f$ of the wavenumber $k$, where $k_0$ is the carrier wavenumber in the system. For the mathematical description of the problem, we start with the transfer matrices $\hat{M}_\mathrm{S}(k)$ and $M_{1,2}(k)$ for the scatterer and for the general optical systems preceding and following the scatterer, respectively. Explicit forms for such matrices are given in Ref.~\cite{Xuereb2009b}. We {then calculate} the transfer matrix $\hat{M}(k)$ of the entire system given by the product $\hat{M}(k)=M_1(k)\,\hat{M}_\mathrm{S}(k)\,M_2(k)${, and} the matrix inverse $M_1^{-1}(k)${, such} that
\begin{equation*}
\begin{pmatrix}
A_\mathrm{l}\\
B_\mathrm{l}
\end{pmatrix}=\hat{M}
\begin{pmatrix}
C_\mathrm{r}\\
D_\mathrm{r}
\end{pmatrix}\,\text{and }
\begin{pmatrix}
A\\
B
\end{pmatrix}=M_1^{-1}\begin{pmatrix}
A_\mathrm{l}\\
B_\mathrm{l}
\end{pmatrix}\,,
\end{equation*}
where we have omitted the $k$-dependence. We use the hat to indicate that the corresponding matrix contains the Doppler shift operator $\hat P_v$. The elements of these matrices, which we {denote, for convenience,} by
\begin{equation}
\label{eq:ABCD}
M_1^{-1} \equiv \bigl[\theta_{ij}\bigr]\text{, and }
\hat{M} \equiv \begin{bmatrix}
\hat{\gamma} & \hat{\alpha}\\
\hat{\delta} & \hat{\beta}
\end{bmatrix}\,,
\end{equation}
can all be obtained in a straightforward manner using only 2-by-2 matrix multiplication for an arbitrary number of scatterers, and hence can in principle be calculated analytically, or can be derived by using formal computer languages.
\par
In order to make the mathematics more concise, we explicitly {consider the case where we pump the system from only one direction}. Setting $C_\mathrm{r}(k)=0$ in \eref{eq:ABCD}, we obtain $A_\mathrm{l}(k)=\hat{\alpha}\hat{\beta}^{-1}B_\mathrm{l}(k)$. Because of the {presence of $\hat{\beta}^{-1}$}, this relation between the back-reflected and the incoming fields contains the powers of the shift operator $\hat{P}_v$ to all orders. In the simple example of one mirror moving in front of a fixed one, the corresponding summation could be carried out analytically~\cite{Xuereb2009b}. However, this is not the case generally. The crucial step to overcome this problem is to express the Doppler shift operator in the transfer matrix $\hat{M}_\mathrm{S}$ to first order in $v/c$: $\hat{P}_v=1+\tfrac{v}{c}k_0\tfrac{\partial}{\partial k}$. Here we have assumed that we pump at a single wavenumber; \ie, we take $B_\mathrm{l}(k)=B_0\,\delta(k-k_0)$, with $\delta(k)$ being the Dirac $\delta$ function and $k_0$ being the wavenumber corresponding to the central pumping frequency. We can thus expand both $\hat{\alpha}$ and $\hat{\beta}$ {in $v/c$ and conveniently denote them by}
\begin{equation*}
\label{eq:BandD}
\hat{\alpha}=\alpha_0+\tfrac{v}{c}\Bigl(\alpha_1^{(0)}+\alpha_1^{(1)}\tfrac{\partial}{\partial k}\Bigr)\,\text{and}\,\hat{\beta}=\beta_0+\tfrac{v}{c}\Bigl(\beta_1^{(0)}+\beta_1^{(1)}\tfrac{\partial}{\partial k}\Bigr)\,.
\end{equation*}
{The auxiliary functions $\alpha_0,\alpha_1^{(0)},\dots$, are simply related to the matrix elements defined in~\eref{eq:ABCD} and to the scattering strength parameter~\cite{Xuereb2009b}, or `polarizability', $\zeta$. We recall that the {amplitude} reflectivity and transmissivity of the scatterer are related to $\zeta$ by $r=i\zeta/(1-i\zeta)$ and $t=1+r$, respectively{; the reflectivity and transmissivity of a mirror or scatterer are, in general, complex and account automatically for phase shifts in the reflected and transmitted fields~\cite{Deutsch1995}.} Thus $\hat{\beta}$ can} be inverted in closed form up to linear order in $v/c$ to yield the amplitude $A(k)$. We then calculate the total field amplitudes $\mathcal{A}=\int A(k)\rmd k$ and $\mathcal{B}=\int B(k)\rmd k$ and {obtain}
\begin{align}
\label{eq:CurlyA}
\mathcal{A}=\Biggl[&\biggl(\theta_{11}\frac{\alpha_0}{\beta_0}+\theta_{12}\biggr)+\frac{v}{c}\Biggl(\theta_{11}\frac{\alpha_1^{(0)}\beta_0-\alpha_0\beta_1^{(0)}}{\beta_0^2}\nonumber\\&-\frac{1}{\beta_0}\frac{\partial}{\partial k}\theta_{11}\frac{\alpha_1^{(1)}\beta_0-\alpha_0\beta_1^{(1)}}{\beta_0}\Biggr)\Biggr]B_0=\mathcal{A}_0+\tfrac{v}{c}\mathcal{A}_1\,,
\end{align}
and similarly for $\mathcal{B}$.
This general solution for the field amplitudes at the scatterer is one of the main results of this Letter, and can be evaluated for an {arbitrary system.} The amplitudes on the
right side of the moving scatterer can be expressed{, using the elements of $\hat{M}_\text{S}$,} as
\begin{equation}
\begin{split}
\mathcal{C}&=(1-i\zeta)\mathcal{A}-i\zeta\bigl(1-2\tfrac{v}{c}\bigr)\mathcal{B}\ \text{and }\\
\mathcal{D}&=i\zeta\bigl(1+2\tfrac{v}{c}\bigr)\mathcal{A}+(1+i\zeta)\mathcal{B}\,,
\end{split}
\label{eq:CurlyCandD}
\end{equation}
where we have used the explicit form of $\hat{M}_\mathrm{S}$, and where we have defined $\mathcal{C}=\int C(k)\rmd k$ and $\mathcal{D}=\int D(k)\rmd k$. In \erefs{eq:CurlyCandD} we have also {assumed that $\zeta$ is independent of $k$. Upon} using these relations, we obtain an expression for the force acting on the scatterer, from which we can extract the friction force {(see Ref.~\cite{Xuereb2009b} for the details of this derivation)}:
\begin{multline}
\label{eq:FullFriction}
\force=-4\hbar k_0\frac{v}{c} \Bigl[\lvert\zeta\rvert^2\bigl(\lvert\mathcal{A}_0\rvert^2-\lvert\mathcal{B}_0\rvert^2\bigr)\\+\bigl(\lvert\zeta\rvert^2+\im{\zeta}\bigr)\re{\mathcal{A}_0\mathcal{A}_1^\star}-2\im{\zeta}\re{\mathcal{A}_0\mathcal{B}_0^\star}\\+\bigl(\lvert\zeta\rvert^2-\im{\zeta}\bigr)\re{\mathcal{B}_0\mathcal{B}_1^\star}+\im{\zeta}\re{\mathcal{A}_0\mathcal{B}_1^\star}\\+\re{\Bigl(\lvert\zeta\rvert^2+i\re{\zeta}\Bigr)\mathcal{A}_1\mathcal{B}_0^\star}\Bigr]\,.
\end{multline}
All our assumptions---{\ie,} pumping at a single wavenumber, frequency independent polarizability ($\partial\zeta/\partial k=0$), and $C_\mathrm{r}(k)=0$---are simplifying assumptions and can be relaxed. However, this would result in forms for the friction force that are less transparent and amenable to analysis.
We now apply this to the `external cavity cooling' configuration{,~\fref{fig:Models}(b)}. As a reference system for the analysis of the cooling force in this setup, we also consider the `mirror mediated cooling' configuration {(see~\fref{fig:Models}(c))}, which has been {previously discussed}~\cite{Xuereb2009b,Xuereb2009a}, and which is {the optomechanical} cooling scheme used in many experiments~\cite{Metzger2004,Arcizet2006,Gigan2006,Schliesser2008}. Note that in the `external cavity cooling' scheme {with a near mirror of complex transmissivity $t$, the limits of small and large $\lvert t\rvert$ render the situation where the cavity is replaced respectively by the near mirror only or the far mirror only.} For intermediate {$t$ compared {with} the transmissivity of the far mirror, $T$}, the moving scatterer interacts with a field reflected back from the {cavity and} is subject to the interference created by the multiple reflections between the two mirrors. In this text, we {consider in particular} an object having low reflectivity, around $50$\%, which corresponds to a polarizability $\zeta=1$ and is representative of typical experimental conditions~\cite{Metzger2004}. This ensures that a high-finesse resonator cannot be formed between the object and the near mirror, thereby guaranteeing a parameter range where the cavity formed between the immobile mirrors dominates the interaction. For the sake of simplicity, we restrict ourselves to the special case of scatterers that can be characterized by a real polarizability; this is equivalent to assuming that no absorption takes place in the scatterer. Similar results hold when $\zeta$ is not real.
\par
\begin{figure}
\centering
\includegraphics[width=\figwidth]{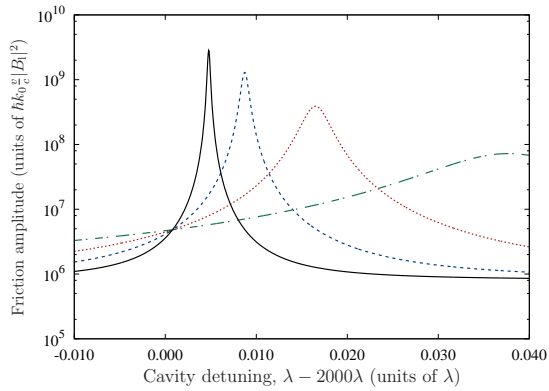}
\caption{The amplitude of the friction force acting on the scatterer, for various near-mirror transmissivities, is shown as a function of the mirror separation in the cavity. The different curves represent different near-mirror transmissivities: $\lvert t\rvert=0.45$ (dashed--dotted curve), $\lvert t\rvert=0.20$ (dotted), $\lvert t\rvert=0.10$ (dashed), $\lvert t\rvert=0.05$ (solid). (Scatterer polarizability $\zeta=1$, scatterer--cavity separation $x\approx 400\lambda_0$, $\lvert T\rvert=0.01$, $\lambda_0=780$~nm.)}
\label{fig:Detuning}
\end{figure}
A numerical fit to \eref{eq:FullFriction} for $\lvert t\rvert\sim\lvert T\rvert$ and $\zeta\ll 1$ renders a friction force of the approximate form
\begin{equation}
\label{eq:ECCOFriction}
\force\approx-8\hbar k_0^2\zeta^2\tfrac{v}{c}(2x+0.17\mathcal{F}L)\sin(4k_0x+\phi)\lvert B_0\rvert^2\,,
\end{equation}
where $\mathcal{F}$ is the cavity finesse, $L$ the cavity length (optimized as discussed below), $x$ the separation between the scatterer and the near mirror, and $\phi$ a phase factor. The gross spatial variation of the friction force is linear in both $L$ and $x$; this is simply because of the linear increase of the retardation time of the reflected field with {the} distance between the scatterer and the mirrors. This dependence is modulated by a wavelength-scale oscillation of the friction force, which thereby follows the same oscillatory dependence as mirror mediated cooling~\cite{Xuereb2009a,Xuereb2009b} and constrains cooling to regions of the size of $\lambda_0/8$, where $\lambda_0=2\pi/k_0$. In the case of a micro-mechanical mirror, where the vibrational amplitude is naturally much less than the wavelength, this presents no problem. The form of \eref{eq:ECCOFriction} is dependent on the properties of the scatterer and of the mirrors; for realistic mirrors and $\zeta=1$, the enhancement factor $0.17\mathcal{F}$ drops to $0.04\mathcal{F}$. With typical experimental parameters this results in an enhancement of $10^3$--$10^4$ over the standard setup; e.g., in Ref.~\cite{Metzger2004} the use of two fixed mirrors could increase the optically induced damping rate, $(\Gamma_\text{eff}-\Gamma)$ in their notation, by over 3 orders of magnitude and lower the limiting temperature from $\Llt 20$\,K to $\Llt 6$\,mK.
\par
As shown in \fref{fig:Detuning}, the fine tuning of the cavity {length by varying $L$ on the wavelength scale shows a Lorentzian-like resonant enhancement} of the friction amplitude, {following that of} the intra-cavity field intensity. If we denote the complex reflectivities of the near and far mirror by $r$ and $R$, respectively, we can show that the peaks of \fref{fig:Detuning} lie around the cavity resonances, at approximately $L=\tfrac{1}{2}m\lambda_0-\tfrac{1}{2k_0}\arg\big(rR\big)$, with $m$ being an integer, and have approximately the same full-width at half-maximum, $\bigl(1-\lvert r R\rvert\bigr)/\bigl(k_0\sqrt{\lvert r R\rvert}\bigr)$. The enhancement of the friction {force by} the cavity {is due to the multiplication of the retardation time by the number of round trips in the cavity, which thereby acts as a `distance folding' mechanism. For the chosen parameters, the optical path length is effectively $2x+0.04\mathcal{F}L$; \ie, determined predominantly by the cavity length $L$.}
\par
\begin{figure}
\centering
\includegraphics[width=\figwidth]{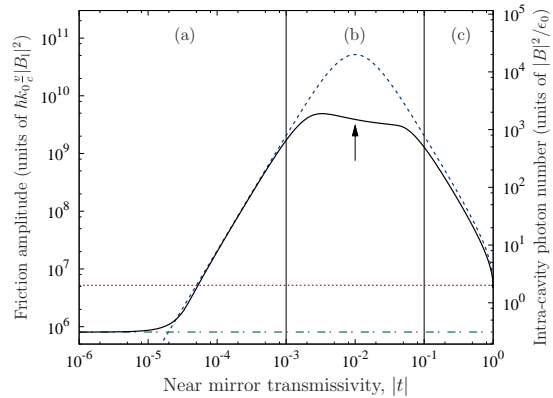}
\caption{Amplitude of the friction acting on a scatterer of polarizability $\zeta=1$ interacting with a cavity tuned to achieve maximum friction, for varying transmissivity of the near mirror. The friction amplitude (solid curve) approaches that for mirror mediated cooling using the far (dotted line, $t\rightarrow 1$) or the near (dashed--dotted line, $t\rightarrow 0$) mirror only in the appropriate limits. The arrow indicates the point at which the two cavity mirrors have the same reflectivity. Also shown is the intra-cavity field (dashed). ($x\approx 400\lambda_0$, $L\approx 2000\lambda_0$, $\lvert T\rvert=0.01$, $\lambda_0=780$~nm, finesse at peak friction $5.0\times 10^4$.)}
\label{fig:Friction}
\end{figure}
\begin{figure}
\centering
\includegraphics[width=\figwidth]{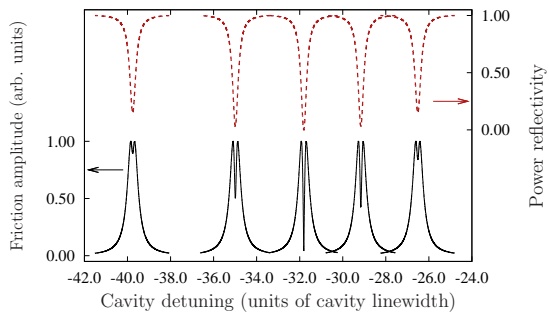}
\caption{In region (b) of \fref{fig:Friction}, the friction coefficient amplitude (solid curves) is attenuated due to the attenuation in the field reflected from the cavity (dashed). $\lvert T\rvert=0.01$ in every plot; $\lvert t\rvert$ is, from left to right, $6.7\times 10^{-3}$, $8.3\times 10^{-3}$, $1.0\times 10^{-2}$, $1.2\times 10^{-2}$, and $1.5\times 10^{-2}$. (Parameters as in~\fref{fig:Friction}.)}
\label{fig:CandR}
\end{figure}

{The friction force depends not only upon the retardation but also upon the cavity reflectivity, which drops near resonance in the well-known behaviour of a Fabry--P\'erot resonator. \fref{fig:Friction} shows the friction amplitude as a function of the near mirror transmissivity $\lvert t \rvert$ for a fixed far mirror transmissivity, $T=1/(1-100i)$.} {We note that this nonideal reflectivity of the far mirror could equivalently arise from absorption, of ca.\ $0.01$\% with the given parameters, of the incident power by the mirror.}
{For each value of $\lvert t \rvert$, the cavity length $L$ has been adjusted to maximize the friction force, according to curves such as those in \fref{fig:Detuning}. The calculated result follows the intra-cavity {field (shown {dashed)} except} where the cavity reflectivity drops near resonance (region (b)), and in the extremes of regions (a) and (c), where the geometry is dominated by the near ($\lvert t\rvert\rightarrow 0$) or far ($\lvert t\rvert\rightarrow 1$) mirrors, respectively. \fref{fig:CandR} shows the effect of the drop in reflectivity as the cavity is scanned through resonance for similar mirror reflectivities. When this causes a dip in the friction amplitude peak, the optimum values plotted in \fref{fig:Friction} occur to either side of the resonance, and the friction force in this region is effectively limited by this interference effect.} {We note that the friction amplitude is not maximized at the point of maximum intra-cavity field ($t=T$)} because more light is lost through the cavity for larger $\lvert t\rvert$.
\par
{{The external cavity cooling mechanism of \fref{fig:Models}(b) may prove particularly valuable when the scatterer is a small mirror or other micro-mechanical optical component. In such cases, the advantage gained by using the external cavity over the standard optomechanical cooling scheme, \fref{fig:Models}(c), depends heavily upon the polarizability or reflectivity of the moving scatterer, which in the above calculations have so far been taken to be modest ($\zeta=1$; $\lvert r\rvert=0.7$) in comparison with those of the cavity mirrors.} For $\zeta\ll 1$, the friction force is enhanced by a factor approximately equal to $\mathcal{F}$} because of the distance folding argument explained above. For larger $\zeta$, the system turns into a three--mirror resonator and the advantage of external cavity cooling is not as big, but is still significant. For $\zeta\approx 1$ we find enhancement by a factor $0.04\mathcal{F}$, as discussed above. For even larger $\zeta$, when the reflectivity of the moving mirror becomes comparable to that of the fixed mirrors, the scheme behaves similarly to the mirror mediated cooling {configuration}. The main heating process that counteracts the cooling effect in the case of micromirrors is thermal coupling to the environment, which depends on the geometry. In the case of isolated scatterers that undergo no absorption, the heating is due to quantum fluctuations in the fields~\cite{Xuereb2009b}; the limit temperature here is $\Lapprox\hbar c/\bigl(0.34k_\text{B}\mathcal{F}L\bigr)$ when $\zeta\ll 1$, which evaluates to $\Lapprox 0.1$\,mK for the parameters in \fref{fig:Friction}.\\
The usual cavity mediated cooling {mechanism}~\cite{Thompson2008,Favero2008}, where the moving scatterer is inside a two-mirror cavity, can also be described by our general framework in terms of \erefs{eq:CurlyA} and (\ref{eq:FullFriction}). Compared with this scheme, external cavity cooling has the advantage of always having a sinusoidal spatial dependence; the narrow resonances in the friction force for well-localized particles in a far-off resonance trap inside a cavity~\cite{vanEnk2001}, for example, impose more stringent positioning requirements. {On the other hand, whereas scatterers travelling distances of many wavelengths within a cavity can experience a net cooling force~\cite{Hechenblaikner1998}, the friction force outside a cavity averages to zero; we find, however, that a net cooling effect arises in a similar geometry in three dimensions which may be particularly significant for micro-mechanical systems~\cite{Horak2010}.} Finally, we note that when the scatterer is outside, rather than within, the cavity the local field is not amplified by the resonator and the incident field can therefore be made much stronger without causing saturation (when the moving scatterer is an atom) or damage (when it is a mirror).\\
\indent{This work was supported by the UK EPSRC (EP/E039839/1 and EP/E058949/1), by the CMMC collaboration within the EuroQUAM programme of the ESF, and by the NSF (NF68736) and NORT (ERC\_HU\_09 OPTOMECH) of Hungary.}

\end{document}